\documentclass[journal=jctcce,manuscript=article]{achemso}
\usepackage[version=3]{mhchem} 
\usepackage{nameref}

\author{Ricardo Manuel Garc\'ia-V\'azquez}
\email{rgarciavazqu@u-bordeaux.fr}

\affiliation[UBordeaux]
{Université de Bordeaux, CNRS UMR 5255, Bordeaux INP, ISM, F-33400 Talence, France}

\author{Lisan David Cabrera-Gonz\'alez}
\affiliation[UManchester]
{Departamento de Física, Facultad de Ciencias, Universidad de Chile, Av. Las Palmeras 3425, Ñuñoa, Santiago, Chile}
\author{Otoniel Denis-Alpizar}
\affiliation[UChile]
{Departamento de Física, Facultad de Ciencias, Universidad de Chile, Av. Las Palmeras 3425, Ñuñoa, Santiago, Chile}
\author{Philippe Halvick}
\affiliation[UBordeaux]
{Université de Bordeaux, CNRS UMR 5255, Bordeaux INP, ISM, F-33400 Talence, France}
\author{Thierry Stoecklin}
\email{thierry.stoecklin@u-bordeaux.fr}
\affiliation[UBordeaux]
{Université de Bordeaux, CNRS UMR 5255, Bordeaux INP, ISM, F-33400 Talence, France}
\title[Extending R--Matrix Theory]
  {Applying R--Matrix Theory to Atom–Molecule Inelastic Collisions: the case study of
\ce{H2O + H}}

\abbreviations{IR,NMR,UV}

\begin{document}


\begin{abstract}
  The present study presents a comprehensive theoretical investigation of atom and asymmetric top molecule inelastic scattering based on the R--matrix formalism. The proposed methodology establishes a rigorous framework for treating inelastic collisions in the space-fixed coordinate system. The excellent numerical performance of the method is demonstrated through the comparison of state-to-state rotationally inelastic R--matrix cross sections for the \ce{H + H2O} system with those obtained using conventional close-coupling (CC) theory. The R-matrix approach is shown to deliver results of comparable accuracy while achieving substantially reduced computation times. The method is furthermore shown to achieve more than one order-of-magnitude speedup by exploiting GPU-accelerated diagonalisation through the MAGMA library. This combination of accuracy and computational efficiency positions the R--matrix approach as a powerful and scalable tool for investigating inelastic scattering involving complex polyatomic systems, thereby paving the way for systematic studies of molecule–molecule interactions in astrophysical, atmospheric, and cold-matter environments.
\end{abstract}

\section{Introduction}
Inelastic collisions between atoms and molecules or between two molecules play a central role in a wide range of environments, from interstellar\cite{Draine2011}to circumstellar media\cite{Massalkhi_2020}, planetary atmospheres\cite{Schunk2009}, plasmas\cite{Capitelli2000} and combustion systems\cite{Law2006}. Accurate quantum mechanical data for such processes are therefore essential in many fields. The prevailing state of the art for modelling inelastic scattering is the close-coupling (CC) method  \cite{arthurs:60}, which treats all the degrees of freedom by solving the Schrödinger equation of the system. While the CC method offers highly accurate results and has been validated with state-of-the-art experiments, its computational cost becomes prohibitively high as the numbers of molecular degrees of freedom and scattering channels increase. Approximations such as the coupled states (CS)\cite{RTPack-CS}, infinite order sudden approximation  (IOSA)\cite{IOSA-Parker-1977} or near neighbours CS (NNCS) \cite{DYang_NNCS} alleviate the cost but often fail to describe resonance features of threshold effects with sufficient accuracy.

Originally developed in the middle of the twentieth century to address nuclear physics collision problems \cite{Wigner1946, Wigner1947}, the R--Matrix theory offers an appealing alternative framework for dealing with such issues. Later, it was very successfully extended outside nuclear physics to a large class of problems involving electron collision with atoms \cite{e-atoms-Burke_1971} or molecules \cite{e-molecule-TENNYSON-2010}, which became its favorite playground. The fundamental concept of the method is based on the division of the molecular configuration space into three regions: an inner region comprising all short-range interactions, an outer region where long-range interactions predominate and where propagation to the asymptotic region can be conducted efficiently, and a final asymptotic region where the interaction potential is negligible and the matching to asymptotic boundary conditions can be performed to extract all the physical information of the collision. 

The R--Matrix theory \cite{Burke-R-Matrix-book} is distinguished by its ability to convert a scattering problem into a calculation of molecular states enclosed in an (hyper)spherical box of finite radius.  This enables the highly efficient tools developed by the electronic structure community over many years to be used to solve the latter problem independently of the collision energy. The ratio of the scattering wave function divided by its derivative (ie the R matrix) at the inner region boundary is then obtained at any collision energy from these box-state energies and wave functions  in a very simple and extremely fast way. It  is then propagated into the outer region up to the asymptotic boundary by using any of the usual R--Matrix propagators\cite{Light1979,BALUJA1984}. Alternatively, it can be inverted to obtain the log-derivative matrix at the boundary and propagated using the log-derivative propagators \cite{johnson:79,manolopoulos:86,manolopoulos:88thesis}. 
This propagation is necessary in the case of electron-molecule collisions since the charge-dipole (or charge-induced dipole) interaction generates a very long range potential. In the present case of collisions between heavy neutral colliders, the long range potential is decreasing a lot faster and we will see that the R--Matrix calculated at the boundary of the inner region allows obtaining $\mathbf{K}$ matrices undistinguishable from the CC one, thus circumventing the propagation in the outer region.  

The R--Matrix theory adds a Bloch operator \cite{Bloch1957} to the molecular Hamiltonian. The sum of both operators is hermitian in the box while it is not the case of the Hamiltonian alone. Furthermore, this enforces the continuity of the wave function derivative at the boundary of the box. This effectively transforms a scattering propagation problem into an eigenvalue problem, a class of problems that has been studied intensively in quantum chemistry for over fifty years, leading to highly optimized and efficient numerical methods.
Solving the inner-region problem is still computationally demanding, both in terms of CPU time and memory usage. However, it only needs to be performed once for each value of the total angular momentum and a given symmetry of the system (parity, exchange of identical particles....).  A further advantage is that GPU-accelerated algebra libraries, such as the NVIDIA Math Libraries\cite{gpu_math} or MAGMA \cite{MAGMA2010}, can be employed to handle large diagonalization tasks efficiently.

While the R--Matrix theory has been successfully used in the field of electron-atom/molecule collisions \cite{Burke2011, Tennyson2010} and nuclear physics \cite{Baye1998, Descouvemont2010}, only a few attempts have been made to apply it to atomic and molecular collisions. Pioneering work was conducted in the 1980s, when \citeauthor{Bocchetta1985} applied the method to a 2D model atom-diatom collision problem \cite{Bocchetta1985} and later to an atom-asymmetric top collision problem\cite{Bocchetta1988}.  Unfortunately, these first studies did not receive much attention within the dynamicist community, most likely because memory was so expensive at that time. The latter study used a very simple model potential, was constrained to a limited energy interval and solely 10 partial waves. The conclusion drawn was that the R-matrix method was incapable of reproducing certain CC resonances  because of the strong limitations of the computer facilities at that time. The Tennyson group recently made a new attempt to popularise the method within the ultra cold collisions community\cite{Tennyson2016} by demonstrating its application to elastic Ar--Ar scattering, first with a Morse potential~\cite{Rivlin_morse_rmat} and later using several \textit{ab initio} potential energy curves~\cite{Rivlin2019}.  The same team then proposed a roadmap for extending its application domain to multiple physical processes in atom-diatom collisions \cite{McKemmish2019}. 

In this study, we present a comprehensive application of R--Matrix theory to inelastic collisions involving an atom and a rotating asymmetric top (RAST) molecule, as exemplified by the collision between atomic hydrogen and rigid water molecule. This system is an ideal benchmark thanks to its astrophysical and atmospheric importance, and to the wealth of previous theoretical studies based on CC method. It furthermore offers a rich physics arising from the strong anisotropy of the  \ce{H + H2O} interaction potential and the large number of accessible rotational states of \ce{H2O}. It is also typical of systems that require the rotational basis set to be limited, as well as other degrees of freedom such as the vibrational bending motion, and to neglect remaining degrees of freedom, such as the vibrational stretching,  in order to avoid an overwhelming increase of the size of the CC calculations. 

In the \nameref{meth} section, we first provide details of how the R--Matrix approach is implemented to treat this system.  In the \nameref{calc} section, we detail the parameters used for the R--Matrix calculations.  In the  \nameref{res&disc} section, the \ce{H + H2O} box-state energies are computed using three different Lagrange Mesh basis of the radial coordinate and compared. The R--Matrix method ability to describe intricate resonance patterns and thresholds is then investigated by comparing with available CC results. The computer time performances of the two methods are also compared and the R--Matrix scalability is eventually discussed in the \nameref{concl}
section.

\section{Methods}
\label{meth}
\subsection{R--Matrix Theory}
Two different versions of the R--Matrix exist and have been widely used in the fields of nuclear, atomic, and molecular physics\cite{Descouvemont2010}. The first one, known as the phenomenological R--Matrix, aims to parametrise  scattering data such as resonance widths and energies. The second one, known as the calculable R--Matrix, which is the one used in this work, aims to accurately solve the Schr\"odinger equation, mainly for scattering problems, but it can be also useful for bound state calculations in the case of very shallow potential well and in the case of a very long-range potential for the states just below the dissociating limit. 

To introduce the definition and derivation of the calculable R--Matrix
theory, let's start by defining the Hamiltonian for the system under study. The Hamiltonian in atomic units for an atom-RAST system in the space-fixed frame is given by: 
\begin{equation}
    \hat{H} = -\dfrac{1}{2\mu} \left(\dfrac{d^2}{dR^2}-\dfrac{l(l+1)}{R^2}\right) + \hat{H}_{RAST} + \hat{V}(\vec{R}),
\end{equation}
where $\mu$ represents the reduced mass, $l(l+1)$ are the eigenvalues of the orbital angular momentum squared operator $\hat{l}^2$, $\hat{V}(\vec{R})$ is the potential energy surface (PES), and $\hat{H}_{RAST}$ is the internal Hamiltonian of an asymmetric top molecule satisfying \cite{zare:88,Garcia-Vazquez_2023,Garcia-Vazquez-2024-FD}: 
\begin{equation}
\hat{H}_{RAST}\left|j\tau m,s\right> = \varepsilon_{j\tau}^{s} \left|j\tau m,s\right>,
\end{equation} 
where $\varepsilon_{j\tau}^{s}$ denote the eigenenergies and $s=0,1$ the parity of the RAST molecule. The total wave function at a given energy $E$ for a given value of the total angular momentum $J$, its projection $M$ to the space-fixed $z$-axis, and the total inversion parity $p$ is given by:
\begin{equation}
    \Psi^{JM p} = \dfrac{1}{R}\sum_{c}  G^{JM p}_{c}(R) \left|cJM \right>,
\end{equation}
where the channel subscript $c$  brings together the four subscripts $j,\tau,s,l$ and $\left|cJM \right>=\left|j\tau s lJM \right>$ represent the channel wavefunctions (we will drop from now on the mentions to $J$,$M$ and $p$ to simplify the notation). The latter wavefunctions are obtained by coupling the angular momentum of the RAST with the relative orbital angular momentum of the molecule as:
\begin{equation}
    \left|c JM \right> = \sum_{m,m_l} \left< j m, l m_l|J M \right > \left|j\tau m,s\right> \left|l m_l\right>,
\end{equation}
Using the expansion of the total wave function in the angular basis, the Schr\"odinger equation for the atom plus molecule system  can be written as a set of coupled equations, usually known as the close-coupling equations:
\begin{equation}
    \sum_{c'} \left [\left(T_c + \varepsilon_{c}- E \right)\delta_{c,c'}  + V_{c,c'}\right]G_{c'}(R) = 0, 
\end{equation}
where 
\begin{equation}
    T_c = -\dfrac{1}{2\mu}\dfrac{d^2}{dR^2}+ \dfrac{l_{c}(l_{c}+1)}{2\mu R^2},
\end{equation}
and with $V_{c,c'}$ representing the matrix elements of the potential,
\begin{equation}
    V_{c,c'} = \left< c\right|\hat{V}\left|c'\right>. 
\end{equation}
Restricting the semi-infinite radial domain of definition of this equation, $R\in [0,+\infty$[, to the finite inner region ($0 < R < a$) makes the Hamiltonian operator non-Hermitian. Adding a Bloch operator \cite{Bloch1957}, 
\begin{equation}
\mathcal{L} = \sum_{c}\left|c\right>\mathcal{L}_{c}\left<c\right|,
\end{equation}
enables to recover the Hermiticity:
\begin{eqnarray}
\sum_{c'} \left [\left(T_c + \varepsilon_{c} - E+\mathcal{L}_{c}\right)\delta_{c,c'}  + V_{c,c'}\right]G_{c'}(R) = \mathcal{L}_{c}G_{c}(R)\label{eq:SBE},\\
\mathcal{L}_{c} = \dfrac{1}{2\mu} \delta(R-a)\left(\dfrac{d}{dR}-\dfrac{B_{c}} {R}\right),\hspace{2.5cm}
\end{eqnarray}
where $B_{c}$ is a real and energy-independent boundary parameter, which is usually assumed to be zero for scattering calculations.

To solve Eqn. \ref{eq:SBE}, the radial wave function in the internal region is expanded in a basis set of $N$ Lagrange mesh functions \cite{Baye1998} $\varphi_i(R)$ as:
\begin{equation}
    G_{c} (R) = \sum_{i=1}^{N} A_{i} \varphi_i(R).
\end{equation}
By using this expansion and defining a $\mathbf{C}$ matrix such as: 
\begin{equation}
    C_{c i \: ; \: c'j \:}(E,B) = \left<\varphi_i\right|\left(T_c +\varepsilon_{c} - E +\mathcal{L}_{c}\right)\delta_{c,c'}  + V_{c,c'}(R)\left|\varphi_j\right>,
\end{equation}
 and after carefully following the methodological development of \citeauthor{Bocchetta1985}\cite{Bocchetta1985} we obtain the expression for the R--Matrix at the boundary as:
\begin{equation}
    R_{c,c'}(E,B) = \dfrac{1}{2\mu a} \sum_{i,j} \varphi_i(a)\left(\mathbf{C}(E,B)\right)^{-1}_{c i,c' j}\varphi_j(a).
\end{equation}
Another definition of the R--Matrix can be given by considering the eigenvalues $E_{n} $ and the corresponding normalised eigenvectors $\boldsymbol{\upsilon}_n$ of the matrix $\mathbf{C}(0,B)$:
\begin{equation}
    \mathbf{C}(0,B)\boldsymbol{\upsilon}_{n} = E_{n} \boldsymbol{\upsilon}_{n},
\end{equation}
and  $\boldsymbol{\upsilon}_n^T\boldsymbol{\upsilon}_{n'} = \delta_{n,n'}$.
Using the eigenvalues and the eigenvectors of the matrix $\mathbf{C}(0,B)$ we can construct the spectral decomposition of the inverse of $\mathbf{C}(E,B)$ as:
\begin{equation}
    [\mathbf{C}(E,B)]^{-1} = \sum_{n} \dfrac{\boldsymbol{\upsilon}_n\boldsymbol{\upsilon}_n^{T}}{E_n-E}\ ,
\end{equation}
and the R--Matrix becomes:
\begin{equation}
    R_{c,c'} (E,B) = \sum_n \dfrac{\gamma_{c n}\gamma_{c' n}}{E_{n}-E}\ ,  
\end{equation}
where the
\begin{equation}
\gamma_{c n} = \left(\dfrac{1}{2\mu a}\right)^{1/2}\sum_i \upsilon_{n,ci} \varphi_i(a),
\end{equation}
are the reduced widths or surface amplitudes which can be also represented as follows:
\begin{equation}
    \gamma_{cn} = \left<c JM\right|\Psi^{JMp} \left.\right>'_{R=a},
\end{equation}
where the prime indicates that the integration is carried out over all variables except $R$.



\subsection{\ce{H2O + H } PES}
The 4D PES used in this work is the one developed by some of the authors and used to study the rotational and vibrational bending relaxation of \ce{H2O} colliding with H \cite{lisan-h-h2o} (P22 PES). A new 6D PES was also developed very recently for this system \cite{H-H2O_6D} (P24 PES), whose accuracy was reported to be similar to that of the P22 PES. In order to compare R--Matrix results with the published CC results obtained using the P22 PES, we perform our R--Matrix calculations using the P22 PES. This allows us to overcome differences arising from the use of different PES.

\section{Calculations}
\label{calc}
Following the R--Matrix approach, we have conducted a study of the relaxation of all the excited rotational states of the water molecule within the $j=1,2$ and $3$ multiplets, leading to 14 initial rotational states. Our basis set contains all the rotational states corresponding to $0\le j \le 9$. Calculations were carried out separately for the  \textit{para} (p) and \textit{ortho} (o) symmetries of the water molecule, each involving a basis set of 50 rotational states.

The inner region equations were solved in the [3, 30] $a_0$ radial interval for 31 values of the total angular momentum ($0 \le J \le 30$), a given total parity of the system (p = 0, 1) and a given p/o symmetry, leading to 124 matrices that needed to be stored.  $N \times N$ Hamiltonian matrices are diagonalised with $N$ ranging from minimum of $N=4000$ to a maximum of $N=67000$. A new code called \texttt{RQMOLS} was developed to solve the inner-part eigenvalue problem and store the eigenenergies and surface amplitudes in external files. By specifying different input options the code is also able to read the stored external files and apply the boundary condition to extract the scattering matrix $\mathbf{S}$ at the boundary of the inner region. As \texttt{RQMOLS} is interfaced with the \texttt{Newmat} code\cite{Stoecklin_h2o-ph2_2019,Stoecklin_he-h2o} the $\mathbf{S}$ matrix can also be calculated inside the \texttt{Newmat} code  at the boundary of the inner region or after propagation of the R--Matrix. 

\section{Results and discussion}
\label{res&disc}
\subsection{Bound States}
We tested three different radial basis sets to solve the Schrödinger-Bloch equation (SBE) in the internal region. They are associated with a Lagrange-Legendre mesh shifted in the [0,1] interval regularised by $R$ denoted LG in Table \ref{boundst} (Eqs. 3.128 and 3.129 Ref. \citenum{BAYE20151}), and two Lagrange-Jacobi meshes shifted in the [0,1] interval regularised by $R$ (Eqs. 3.178 and 3.179 Ref. \citenum{BAYE20151}) with $a=0,b=3$ denoted JC1 and $a=0,b=30$ denoted JC2. The results obtained for the low-lying bound states of the system are compared with the eigenvalues of the Schrödinger equation obtained with a 200-point Chebyshev DVR (CH)\cite{Stoecklin_he-h2o,Garcia-Vazquez_2023,Garcia_Vazquez_2024_ar} in Table \ref{boundst}. The computed eigenenergies  of the Schrödinger and Schrödinger-Bloch equations are seen to be in excellent agreement. The different Lagrange meshes also produce practically the same results. For this reason, we opted for the Lagrange-Legendre mesh for all the calculations.

While for the present bound state calculations,  the  $B_{\alpha} = 0$ choice ensures an excellent agreement between the Schrödinger and Schrödinger-Bloch eigenvalues, it is worth mentioning  that it may not be the case for very specific problems. This point was discussed, for example by \citeauthor{Baye1998}\cite{Baye1998}, where to ensure a good agreement between the calculated and analytically known bound state energies, $B_{\alpha}$ needed to be optimised in an iterative fashion. In the next paragraph which is dedicated to inelastic scattering calculations, since the choice of $B_{\alpha}=0$ has been proposed as the optimal one in the literature, this is the choice that we will use.
\begin{table}[!ht]
    \centering
    \begin{tabular}{ccrrr}
    \hline 
    \hline 
    Assignment & Radial basis & \textbf{$J=0$}	& \textbf{$J=1$} & \textbf{$J=2$}  \\
    \hline 
    \ce{H-pH2O} & & & & \\
    $\Sigma (0_{00})^e$		& CH  & $-12.677$ & $-10.372$ & $-5.914$   \\
                        & LG  & $-12.676$ & $-10.371$ & $-5.910$   \\
                        & JC1 & $-12.676$ & $-10.371$ & $-5.910$   \\
                        & JC2 & $-12.676$ & $-10.371$ & $-5.910$   \\
                        
$\Pi (1_{11})^e$		& CH  &           & $ 24.189$ & $ 26.479$  \\
                        & LG  &           & $ 24.249$ & $ 26.295$  \\
                        & JC1 &           & $ 24.249$ & $ 26.295$  \\
                        & JC2 &           & $ 24.249$ & $ 26.295$  \\

$\Pi (1_{11})^f$		& CH  &           & $ 27.031$ & $ 31.471$  \\
                        & LG  &           & $ 27.032$ & $ 31.476$  \\
                        & JC1 &           & $ 27.032$ & $ 31.476$  \\
                        & JC2 &           & $ 27.032$ & $ 31.476$  \\
\hline 
\ce{H-oH2O} & & & & \\
$\Sigma (1_{01})^e$		& CH  & $12.622$  & $11.066$  & $13.311$   \\
                        & LG  & $12.623$  & $11.067$  & $13.313$   \\
                        & JC1 & $12.623$  & $11.067$  & $13.313$   \\
                        & JC2 & $12.623$  & $11.067$  & $13.313$   \\
                        
$\Pi (1_{10})^e$		& CH  &           & $ 17.540$ & $ 23.780$  \\
                        & LG  &           & $ 17.544$ & $ 23.814$  \\
                        & JC1 &           & $ 17.544$ & $ 23.814$  \\
                        & JC2 &           & $ 17.544$ & $ 23.814$  \\

$\Pi (1_{10})^f$		& CH  &           & $ 13.814$ & $ 18.251$  \\
                        & LG  &           & $ 13.815$ & $ 18.257$  \\
                        & JC1 &           & $ 13.814$ & $ 18.257$  \\
                        & JC2 &           & $ 27.815$ & $ 18.257$  \\
\hline 
\hline 
\end{tabular}
    \caption{Low lying bound states of the \ce{H-H2O} system. CH denotes the eigenvalues of the Schrödinger equation obtained using a Chebyshev DVR grid. LG, JC1 and JC2 denote respectively the eigenvalues of the Schrödinger-Bloch equation obtained using a Shifted Lagrange- Legendre mesh regularised by $R$, a shifted Lagrange-Jacobi mesh regularised by $R$ with $\beta=3$ and a shifted Lagrange-Jacobi mesh regularised by $R$ with $\beta=30$. }
    \label{boundst}
\end{table}

\subsection{Rotationally inelastic cross sections}
Cross sections for relaxation from all excited rotational states belonging to the $j = 1, 2$ and $3 $ multiplets of the water molecule have been computed for comparison with previous CC  results. \cite{lisan-h-h2o}. These multiplets correspond to $7$ \textit{para} and $7$ \textit{ortho} states. In our previous study, the CC calculations used a log-derivative propagator over the [3-30]$a_0$ interval with a constant step size of 0.1 $a_0$. The current R--Matrix calculations are then performed in the same  internal region from [3-30]$a_0$ and use a 200-point LG basis. The boundary conditions are applied directly at $R = a = 30$ $a_0$ without any further propagation being performed. We include partial waves up to $J_{max}=30$ ensuring a relative convergence of the cross sections better than $10^{-3}$. 

\begin{figure}[H]
\includegraphics[width=\linewidth]{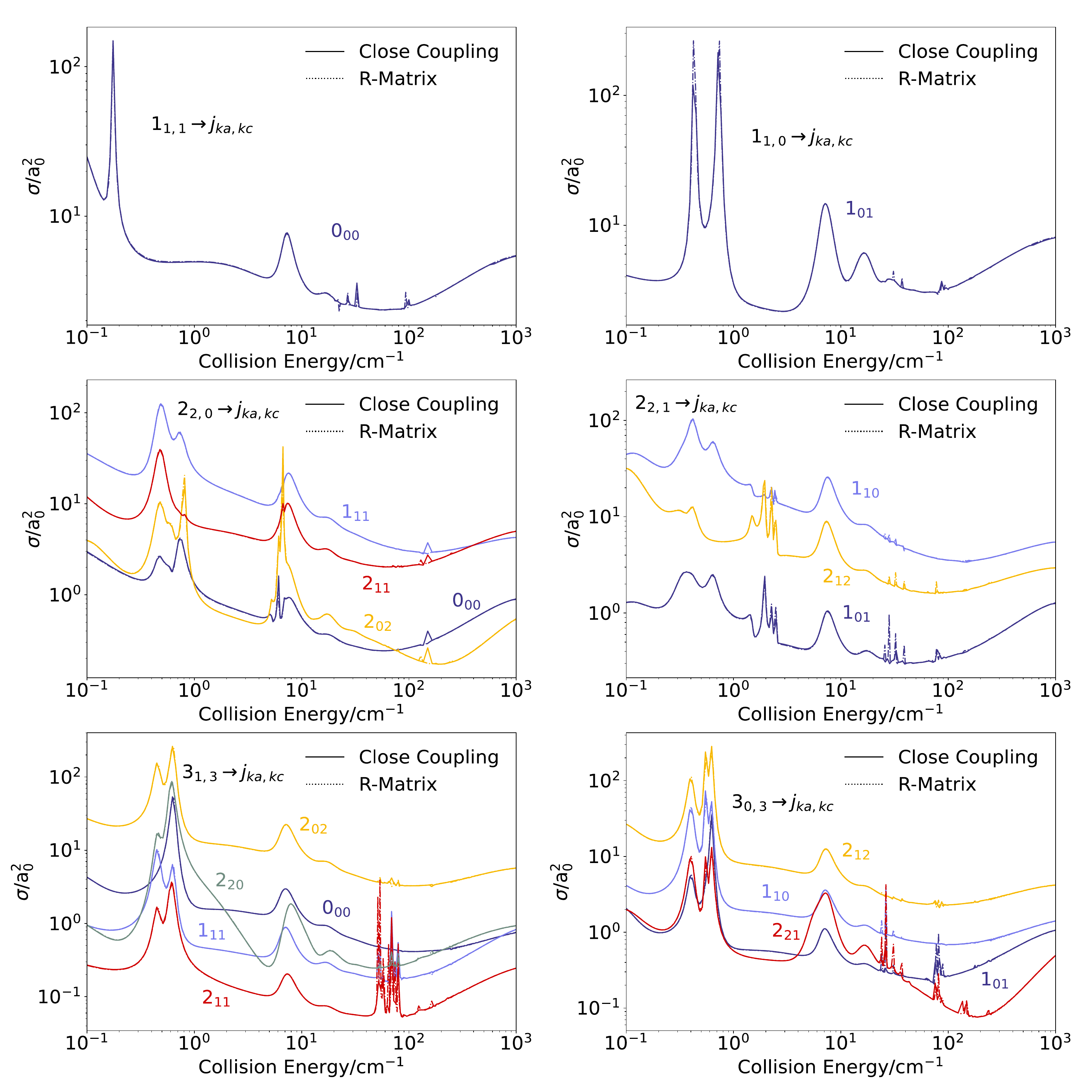}
\caption{State-to-state cross sections for the rotational relaxation of \ce{H2O} by collision with \ce{H} from several \textit{para/ortho} initial states. Solid lines represent the CC cross sections and dotted lines the R--Matrix cross-sections. \label{cross} Solid and dotted lines are fully overlapping, except for fine resonances structures.}
\end{figure}

The results obtained for the \textit{ortho} $1_{10},2_{21}, 3_{03}$ and \textit{para}  $1_{11},2_{20}, 3_{13}$ initial states of \ce{H2O} are presented in Figure \ref{cross}. The R--Matrix and CC calculations show a high level of agreement, with the two types of curve being almost indistinguishable regardless of the initial state considered. The only minor discrepancies observed in the most pronounced resonant peaks can be attributed to the use of different energy grids. This level of accuracy is to be expected given that the R--Matrix method is formally exact and the same approximations are used for both types of calculation. 

This comparison also shows that applying the asymptotic boundary conditions at the boundary of the inner region produces accurate results, eliminating the need for further propagation. This is true of the type of system considered in this study, namely a collision between neutral particles which involves usually an interaction potential that decay sufficiently rapidly. This would not be true for systems involving stronger long-range potentials, such as ion-neutral or ion-ion collisions, or in the ultracold regime where the asymptotic matching radius is so large that direct diagonalization over the entire interval becomes computationally impractical. 

\begin{figure}[H]
\includegraphics[width=0.8\linewidth]{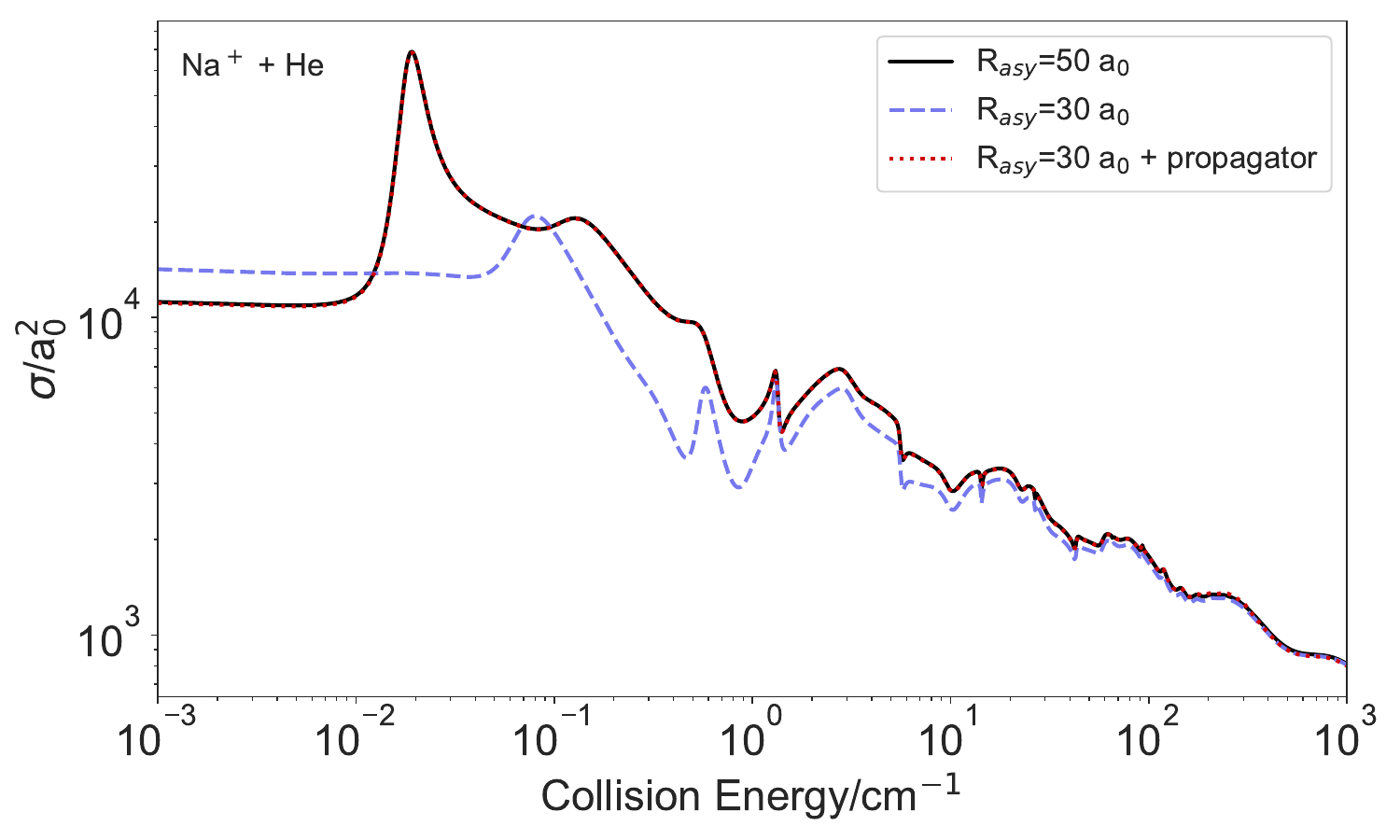}
\caption{ Elastic cross section for the Na$^{+}$ + He collision. Black solid lines represent the results obtained using the R--Matrix method in a $[0.2,50]\,a_0$ interval, dashed blue lines represent the R--Matrix calculations restricted to the shorter interval $[0.2,30]\,a_0$ without propagation while dotted red lines represent calculations over $[0.2,30]\,a_0$ supplemented by propagation from $30$ to $50\,a_0$. The red dotted and black solid lines are fully overlapping\label{elast}}
\end{figure}  

To illustrate this point, we will consider the simple example of an elastic collision between an ion and an atom, namely the Na$^+$ + He system. The potential energy curve (PEC) used here is the same than the one used in a previous study.~\cite{GarciaVazquez2025}. In this one-dimensional scenario, increasing the number of grid points within the internal region does not present any computational challenges. Nevertheless, to demonstrate the role of R--Matrix propagation, we present in Fig.~\ref{elast} three different scenarios: (i) R--Matrix calculations without propagation over the interval $[0.2,50]\,a_0$; (ii) calculations restricted to the shorter interval $[0.2,30]\,a_0$ without propagation; and (iii) calculations over $[0.2,30]\,a_0$ supplemented by propagation from $30$ to $50\,a_0$ using the BBM R--Matrix propagator~\cite{BALUJA1984}. As can be seen, the calculation restricted to the shorter interval only converges for collision energies above 100~cm$^{-1}$ and fails to reproduce the cross section at lower energies. By contrast, performing a short propagation over the interval  $[30,50]\,a_0$ restores convergence with respect to the asymptotic matching radius, producing results from cases (i) and (iii) that are practically indistinguishable.

\subsubsection{Computational performance}
\begin{figure}[H]
\includegraphics[width=\linewidth]{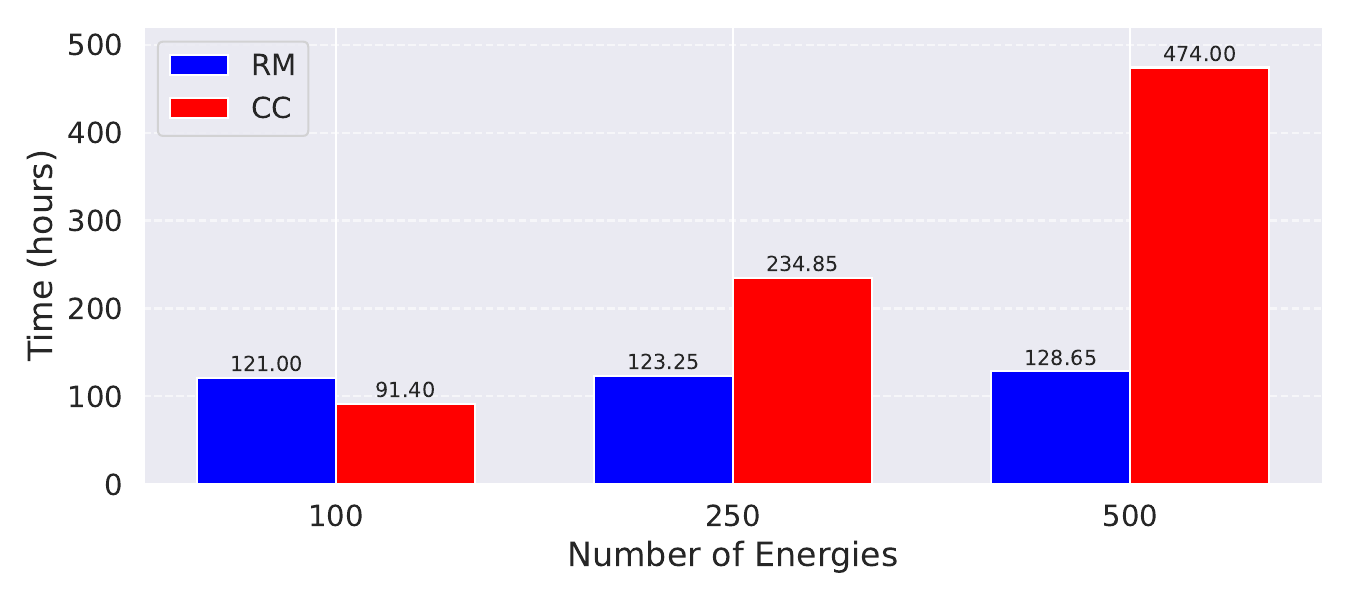}
\caption{Comparison of the R--Matrix ( in blue ) and  CC (in red) computation time. We consider three grids of respectively 100, 250, and 500 energies in the  [0.1,1000] cm$^{-1}$ interval. The computation time required is specified over each bar. \label{time}}
\end{figure}

We conducted a test comparing the total computer time required by the two methods. The first step of the R--Matrix approach which involves the construction and the  diagonalization of 124 matrices needs 240 h and 24 min on 32 cores of the Intel Cascade Lake 6248 processor, with computation time for the \textit{para} or \textit{ortho} species being almost the same. Given that the computation were carried out independently for p-\ce{H2O} and o-\ce{H2O}, the comparison between the R--Matrix and the CC computation times will be conducted solely for collisions involving one of the two species, the p-\ce{H2O}. We then compare in Fig. \ref{time} the R--Matrix and CC computation times as a function of the number of collision energies for which the cross section are calculated. The computation time presented is the one required to achieve the partial waves convergence of the cross section for five initial states independently of the size of the rotational basis set of p-\ce{H2O} which is constant.

As illustrated in this figure, for the smaller energy grid studied (100 energy points), the CC performs better than the R--Matrix. This is due to the large time required in the diagonalization in the case of the R--Matrix. Indeed the time required in the R--Matrix calculation for this energy grid correspond to 121 hours, of which 120.20 hours correspond to the diagonalizations. For the two other energy grids studied, R--Matrix dominates the comparison, with times significantly smaller than the respective CC times. This can be explained by realizing that in the R--Matrix calculations, if we exclude the diagonalization time, which is independent of the number of energy, the time required for each grid is 0.80, 3.05, and 8.45 hours respectively. This time is practically negligible with respect to the total time required by the diagonalizations, making the scale in time of the R--Matrix very efficient.  
When considering the convergence of multiple initial states and large grids of collision energies, the computational advantage of the R--Matrix is becoming more and more important. Significant computational gains can therefore be expected when dealing with the inelastic dynamics problems that are currently of interest in the fields of astrochemistry and atmospheric chemistry. The computational advantage of the R--Matrix over CC calculations simply comes from the fact that, for the last method, a large amount of computer time is used to calculate the potential matrix values at each point on the radial grid. In the case of two neutral fragments interacting, we have seen that the asymptotic boundary conditions can be applied at the boundary of the inner region of the R--Matrix method, with no further propagation required. This is a pivotal point that will increase the advantage of the R--Matrix method when considering molecules with a greater number of internal degrees of freedom, as this adds complexity to the calculation of the matrix elements of the potential. 

\begin{figure}[H]
\includegraphics[width=\linewidth]{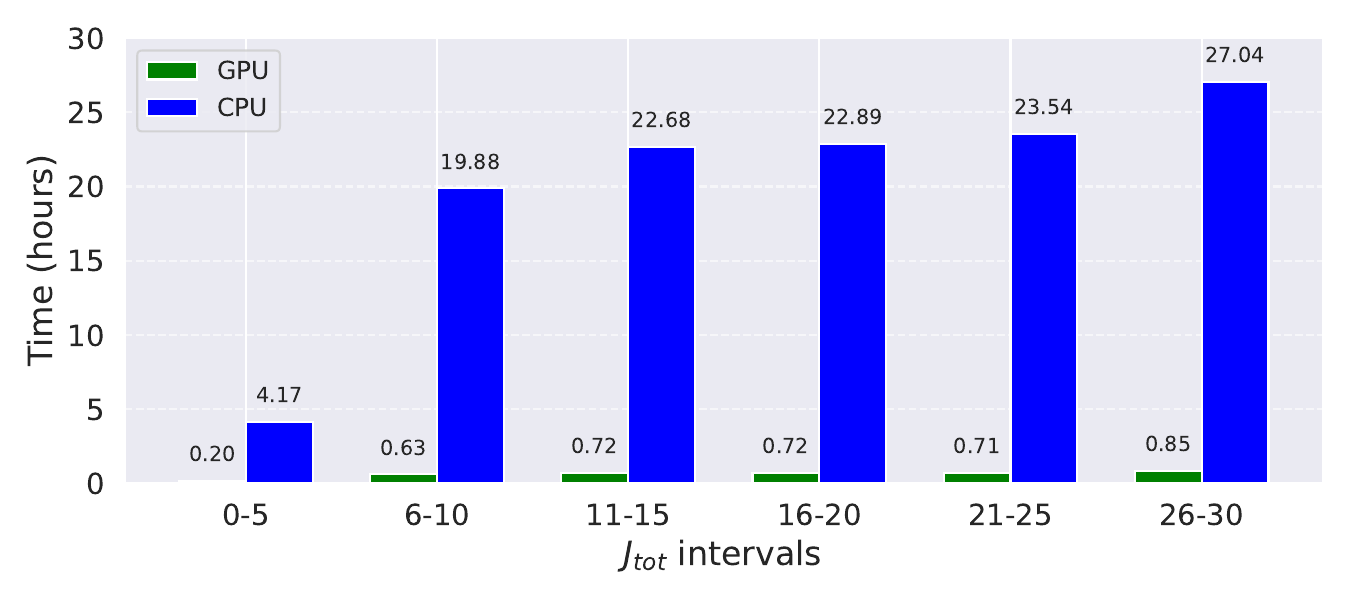}
\caption{Comparison of the R--Matrix time performance using CPU ( in blue ) and  GPUs (in green) to carry out the diagonalizations. We have divided the diagonalizations in intervals of the total angular momentum $J$. For each interval we have solve the diagonalization of all the values included in the interval and the two total parities. \label{time2}}
\end{figure}

Further improvements in the R-matrix computation times can be achieved by reducing the cost of the diagonalization step, which—as discussed previously—represents the most computationally demanding part of the R-matrix workflow. To address this limitation, we explored the use of GPUs to accelerate the diagonalization process. Specifically, we performed the same diagonalization using the MAGMA GPU library \cite{MAGMA2010}, employing the \texttt{dsyevd\_m} routine on three H100 GPUs in the \texttt{gpu\_p6} partition of the Jean Zay supercomputer \cite{jeanzayGPU}. With these hardware specifications, the total diagonalization time decreased from 120.20 hours (CPU execution) to 3.83 hours (GPU execution), resulting in a substantial acceleration of the computational workflow.

A comparison of the diagonalization times for six angular momentum intervals is presented in Fig.~\ref{time2}. For each interval, we have carried out 10 diagonalizations corresponding to the five values of the total angular momentum and the two total parities, except in the first interval where 12 diagonalizations have been carried out. As shown in the figure, the use of GPUs leads to a dramatic reduction in total computation time, highlighting the remarkable efficiency of GPU-accelerated diagonalization. This achievement opens promising perspectives for extending R-matrix calculations to  larger molecular systems.

\section{Conclusions}
\label{concl}
In this work, we presented a comprehensive application of the R--Matrix method to inelastic collisions involving an atom and a RAST molecule, as exemplified by the collision between H and  rigid \ce{H2O}. This system was chosen because its medium size is representative of what is typically needed in astrochemistry. We performed R--Matrix calculations of the inelastic relaxation cross section for all the excited rotational states belonging to the $j$ = 1, 2 and 3 multiplets of the water molecule, and compared the results with the CC calculations presented in a previous study \cite{lisan-h-h2o}. The R--Matrix calculations in the inner region were carried out using a newly developed code \texttt{RQMOLS} interfaced with a modified version of  the \texttt{NEWMAT}\cite{Stoecklin_h2o-ph2_2019,Stoecklin_he-h2o} CC code allowing both to propagate the R matrix and apply the boundary conditions to extract the $\mathbf{S}$ matrix. The agreement between the R--Matrix and CC calculations is found to be remarkable, with the curves representing the R--Matrix and CC state-to-state cross sections being almost indistinguishable. These results demonstrate the R--Matrix method's excellent performance in treating complex inelastic molecular scattering problems. We further tested the two possibilities of extracting the $\mathbf{S}$ matrix directly from the R--Matrix calculated at the boundary of the inner region or after propagation in the asymptotic region. In the present case of interaction between two neutral species, we find that the long-range potential is not strong enough to require propagation over long distances. In other words the S matrix can be extracted  directly from the R--Matrix at the boundary of the inner region, which makes the R--Matrix calculations even faster.  This would not be the case for ion-neutral collisions which would require further propagation as illustrated in the paragraph dedicated to the  Na$^+$ + He collision. We eventually compared the computation times required for the R-matrix and CC calculations and found that the former provides a clear performance advantage. As shown before, for small energy grid, CC offers a better performance. However, the R--matrix method becomes more attractive for larger energy grids, becoming 1.3× faster for the 250 energy grid and up to 3.7× faster for the largest one, highlighting its superior scalability with increasing energy resolution. This advantage becomes even more pronounced when using GPUs, which enable the rapid diagonalization of large matrices. In particular, employing the MAGMA library on the H100 GPUs partition reduces the diagonalization time from 120.20 hours (CPU) to 3.83 hours (GPU),  a remarkable 31-fold speedup that significantly enhances the overall computational efficiency of the R--matrix approach.

In conclusion, the present results establish that the R--Matrix method is a viable and efficient alternative to the CC formalism for modelling inelastic collisions between atoms and molecules. Its scalability,  considered in conjunction with the potential for leveraging contemporary high-performance computing architectures, suggests a broad applicability to more complex systems where CC calculations are currently impractical. Subsequent research will concentrate on extending the methodology to encompass additional degrees of freedom and to address polyatomic systems of astrophysical and atmospheric relevance.

\begin{acknowledgement}
This work was performed using HPC resources from GENCI-IDRIS (Grant 2025-AD010816507 (RMGV) and  Grant 2025-AD010816183 (TS)). Support from the French \textit{Agence Nationale de la Recherche} (\textbf{ANR-Waterstars}), Contract No. ANR-20-CE31-0011, and ANID/FONDECYT Exploración No.13250044 projects is gratefully acknowledged. RMGV thanks Philippe Aurel for a helpful discussion regarding the software optimisation.  We also acknowledge the use of the computing facility of \textit{Universit\'{e} de Bordeaux et Universit\'{e} de Pau et des Pays de l'Adour}.
\end{acknowledgement}




\bibliography{biblio}

\providecommand{\latin}[1]{#1}
\makeatletter
\providecommand{\doi}
  {\begingroup\let\do\@makeother\dospecials
  \catcode`\{=1 \catcode`\}=2 \doi@aux}
\providecommand{\doi@aux}[1]{\endgroup\texttt{#1}}
\makeatother
\providecommand*\mcitethebibliography{\thebibliography}
\csname @ifundefined\endcsname{endmcitethebibliography}  {\let\endmcitethebibliography\endthebibliography}{}
\begin{mcitethebibliography}{44}
\providecommand*\natexlab[1]{#1}
\providecommand*\mciteSetBstSublistMode[1]{}
\providecommand*\mciteSetBstMaxWidthForm[2]{}
\providecommand*\mciteBstWouldAddEndPuncttrue
  {\def\EndOfBibitem{\unskip.}}
\providecommand*\mciteBstWouldAddEndPunctfalse
  {\let\EndOfBibitem\relax}
\providecommand*\mciteSetBstMidEndSepPunct[3]{}
\providecommand*\mciteSetBstSublistLabelBeginEnd[3]{}
\providecommand*\EndOfBibitem{}
\mciteSetBstSublistMode{f}
\mciteSetBstMaxWidthForm{subitem}{(\alph{mcitesubitemcount})}
\mciteSetBstSublistLabelBeginEnd
  {\mcitemaxwidthsubitemform\space}
  {\relax}
  {\relax}

\bibitem[Draine(2011)]{Draine2011}
Draine,~B.~T. \emph{Physics of the Interstellar and Intergalactic Medium}; Princeton University Press: Princeton, NJ, 2011; See Chapter 2: Collisions and Radiative Processes\relax
\mciteBstWouldAddEndPuncttrue
\mciteSetBstMidEndSepPunct{\mcitedefaultmidpunct}
{\mcitedefaultendpunct}{\mcitedefaultseppunct}\relax
\EndOfBibitem
\bibitem[{Massalkhi, S.} \latin{et~al.}(2020){Massalkhi, S.}, {Agúndez, M.}, {Cernicharo, J.}, and {Velilla-Prieto, L.}]{Massalkhi_2020}
{Massalkhi, S.}; {Agúndez, M.}; {Cernicharo, J.}; {Velilla-Prieto, L.} The abundance of S- and Si-bearing molecules in O-rich circumstellar envelopes of AGB stars. \emph{Astronomy \& Astrophysics} \textbf{2020}, \emph{641}, A57\relax
\mciteBstWouldAddEndPuncttrue
\mciteSetBstMidEndSepPunct{\mcitedefaultmidpunct}
{\mcitedefaultendpunct}{\mcitedefaultseppunct}\relax
\EndOfBibitem
\bibitem[Schunk and Nagy(2009)Schunk, and Nagy]{Schunk2009}
Schunk,~R.~W.; Nagy,~A.~F. \emph{Ionospheres: Physics, Plasma Physics, and Chemistry}, 2nd ed.; Cambridge University Press: Cambridge, UK, 2009; Includes detailed discussion of elastic and inelastic collisions in planetary atmospheres\relax
\mciteBstWouldAddEndPuncttrue
\mciteSetBstMidEndSepPunct{\mcitedefaultmidpunct}
{\mcitedefaultendpunct}{\mcitedefaultseppunct}\relax
\EndOfBibitem
\bibitem[Capitelli \latin{et~al.}(2000)Capitelli, Ferreira, Gordiets, and Osipov]{Capitelli2000}
Capitelli,~M.; Ferreira,~C.~M.; Gordiets,~B.~F.; Osipov,~A.~I. \emph{Plasma Kinetics in Atmospheric Gases}; Springer Series on Atomic, Optical, and Plasma Physics; Springer: Berlin, Heidelberg, 2000; Vol.~31\relax
\mciteBstWouldAddEndPuncttrue
\mciteSetBstMidEndSepPunct{\mcitedefaultmidpunct}
{\mcitedefaultendpunct}{\mcitedefaultseppunct}\relax
\EndOfBibitem
\bibitem[Law(2006)]{Law2006}
Law,~C.~K. \emph{Combustion Physics}; Cambridge University Press: Cambridge, UK, 2006; Covers collisional energy transfer, unimolecular falloff, and pressure-dependent reaction kinetics\relax
\mciteBstWouldAddEndPuncttrue
\mciteSetBstMidEndSepPunct{\mcitedefaultmidpunct}
{\mcitedefaultendpunct}{\mcitedefaultseppunct}\relax
\EndOfBibitem
\bibitem[Arthurs and Dalgarno(1960)Arthurs, and Dalgarno]{arthurs:60}
Arthurs,~A.~M.; Dalgarno,~A. The theory of scattering by a rigid rotator. \emph{Proceedings of the Royal Society of London. Series A. Mathematical and Physical Sciences} \textbf{1960}, \emph{256}, 540--551\relax
\mciteBstWouldAddEndPuncttrue
\mciteSetBstMidEndSepPunct{\mcitedefaultmidpunct}
{\mcitedefaultendpunct}{\mcitedefaultseppunct}\relax
\EndOfBibitem
\bibitem[Pack(1974)]{RTPack-CS}
Pack,~R.~T. Space‐fixed vs body‐fixed axes in atom‐diatomic molecule scattering. Sudden approximations. \emph{The Journal of Chemical Physics} \textbf{1974}, \emph{60}, 633--639\relax
\mciteBstWouldAddEndPuncttrue
\mciteSetBstMidEndSepPunct{\mcitedefaultmidpunct}
{\mcitedefaultendpunct}{\mcitedefaultseppunct}\relax
\EndOfBibitem
\bibitem[Parker and Pack(1978)Parker, and Pack]{IOSA-Parker-1977}
Parker,~G.~A.; Pack,~R.~T. Rotationally and vibrationally inelastic scattering in the rotational IOS approximation. Ultrasimple calculation of total (differential, integral, and transport) cross sections for nonspherical molecules. \emph{The Journal of Chemical Physics} \textbf{1978}, \emph{68}, 1585--1601\relax
\mciteBstWouldAddEndPuncttrue
\mciteSetBstMidEndSepPunct{\mcitedefaultmidpunct}
{\mcitedefaultendpunct}{\mcitedefaultseppunct}\relax
\EndOfBibitem
\bibitem[Yang \latin{et~al.}(2018)Yang, Hu, Zhang, and Xie]{DYang_NNCS}
Yang,~D.; Hu,~X.; Zhang,~D.~H.; Xie,~D. An improved coupled-states approximation including the nearest neighbor Coriolis couplings for diatom-diatom inelastic collision. \emph{The Journal of Chemical Physics} \textbf{2018}, \emph{148}, 084101\relax
\mciteBstWouldAddEndPuncttrue
\mciteSetBstMidEndSepPunct{\mcitedefaultmidpunct}
{\mcitedefaultendpunct}{\mcitedefaultseppunct}\relax
\EndOfBibitem
\bibitem[Wigner(1946)]{Wigner1946}
Wigner,~E.~P. Resonance Reactions and Anomalous Scattering. \emph{Physical Review} \textbf{1946}, \emph{70}, 15\relax
\mciteBstWouldAddEndPuncttrue
\mciteSetBstMidEndSepPunct{\mcitedefaultmidpunct}
{\mcitedefaultendpunct}{\mcitedefaultseppunct}\relax
\EndOfBibitem
\bibitem[Wigner and Eisenbud(1947)Wigner, and Eisenbud]{Wigner1947}
Wigner,~E.~P.; Eisenbud,~L. Higher Angular Momenta and Long Range Interaction in Resonance Reactions. \emph{Physical Review} \textbf{1947}, \emph{72}, 29\relax
\mciteBstWouldAddEndPuncttrue
\mciteSetBstMidEndSepPunct{\mcitedefaultmidpunct}
{\mcitedefaultendpunct}{\mcitedefaultseppunct}\relax
\EndOfBibitem
\bibitem[Burke \latin{et~al.}(1971)Burke, Hibbert, and Robb]{e-atoms-Burke_1971}
Burke,~P.~G.; Hibbert,~A.; Robb,~W.~D. Electron scattering by complex atoms. \emph{Journal of Physics B: Atomic and Molecular Physics} \textbf{1971}, \emph{4}, 153\relax
\mciteBstWouldAddEndPuncttrue
\mciteSetBstMidEndSepPunct{\mcitedefaultmidpunct}
{\mcitedefaultendpunct}{\mcitedefaultseppunct}\relax
\EndOfBibitem
\bibitem[Tennyson(2010)]{e-molecule-TENNYSON-2010}
Tennyson,~J. Electron–molecule collision calculations using the R-matrix method. \emph{Physics Reports} \textbf{2010}, \emph{491}, 29--76\relax
\mciteBstWouldAddEndPuncttrue
\mciteSetBstMidEndSepPunct{\mcitedefaultmidpunct}
{\mcitedefaultendpunct}{\mcitedefaultseppunct}\relax
\EndOfBibitem
\bibitem[Burke(2011)]{Burke-R-Matrix-book}
Burke,~P.~G. \emph{R-Matrix Theory of Atomic Collisions}; Springer Series on Atomic, Optical and Plasma Physics; Springer: Berlin, Heidelberg, 2011\relax
\mciteBstWouldAddEndPuncttrue
\mciteSetBstMidEndSepPunct{\mcitedefaultmidpunct}
{\mcitedefaultendpunct}{\mcitedefaultseppunct}\relax
\EndOfBibitem
\bibitem[Light \latin{et~al.}(1979)Light, Walker, Stechel, and Schmalz]{Light1979}
Light,~J.~C.; Walker,~R.~B.; Stechel,~E.~B.; Schmalz,~T.~G. R-matrix propagation methods in inelastic and reactive collisions. \emph{Computer Physics Communications} \textbf{1979}, \emph{17}, 89--97\relax
\mciteBstWouldAddEndPuncttrue
\mciteSetBstMidEndSepPunct{\mcitedefaultmidpunct}
{\mcitedefaultendpunct}{\mcitedefaultseppunct}\relax
\EndOfBibitem
\bibitem[Baluja \latin{et~al.}(1984)Baluja, Burke, and Morgan]{BALUJA1984}
Baluja,~K.; Burke,~P.; Morgan,~L. R-matrix propagation program for solving coupled second-order differential equations. \emph{Computer Physics Communications} \textbf{1984}, \emph{35}, C--823\relax
\mciteBstWouldAddEndPuncttrue
\mciteSetBstMidEndSepPunct{\mcitedefaultmidpunct}
{\mcitedefaultendpunct}{\mcitedefaultseppunct}\relax
\EndOfBibitem
\bibitem[Johnson(1979)]{johnson:79}
Johnson,~B.~R. The log-derivative and renormalized Numerov algorithms. \emph{NRCC Proceedings} \textbf{1979}, \emph{5}, 86\relax
\mciteBstWouldAddEndPuncttrue
\mciteSetBstMidEndSepPunct{\mcitedefaultmidpunct}
{\mcitedefaultendpunct}{\mcitedefaultseppunct}\relax
\EndOfBibitem
\bibitem[Manolopoulos(1986)]{manolopoulos:86}
Manolopoulos,~D.~E. An improved log derivative method for inelastic scattering. \emph{The Journal of Chemical Physics} \textbf{1986}, \emph{85}, 6425--6429\relax
\mciteBstWouldAddEndPuncttrue
\mciteSetBstMidEndSepPunct{\mcitedefaultmidpunct}
{\mcitedefaultendpunct}{\mcitedefaultseppunct}\relax
\EndOfBibitem
\bibitem[Manolopoulos(1988)]{manolopoulos:88thesis}
Manolopoulos,~D.~E. Close-coupling equations the log derivative approach to inelastic scattering, bound states and photofragmentation problems. Ph.D.\ thesis, University of Cambridge, 1988\relax
\mciteBstWouldAddEndPuncttrue
\mciteSetBstMidEndSepPunct{\mcitedefaultmidpunct}
{\mcitedefaultendpunct}{\mcitedefaultseppunct}\relax
\EndOfBibitem
\bibitem[Bloch(1957)]{Bloch1957}
Bloch,~C. Une formulation unifiée de la théorie des réactions nucléaires. \emph{Nuclear Physics} \textbf{1957}, \emph{4}, 503--528\relax
\mciteBstWouldAddEndPuncttrue
\mciteSetBstMidEndSepPunct{\mcitedefaultmidpunct}
{\mcitedefaultendpunct}{\mcitedefaultseppunct}\relax
\EndOfBibitem
\bibitem[Jhunjhunwala and Talavera(2022)Jhunjhunwala, and Talavera]{gpu_math}
Jhunjhunwala,~A.; Talavera,~G. Accelerating {GPU} Applications with {NVIDIA} Math Libraries. NVIDIA Developer Blog, 2022; \url{https://developer.nvidia.com/blog/accelerating-gpu-applications-with-nvidia-math-libraries/}\relax
\mciteBstWouldAddEndPuncttrue
\mciteSetBstMidEndSepPunct{\mcitedefaultmidpunct}
{\mcitedefaultendpunct}{\mcitedefaultseppunct}\relax
\EndOfBibitem
\bibitem[Tomov \latin{et~al.}(2010)Tomov, Dongarra, and Baboulin]{MAGMA2010}
Tomov,~S.; Dongarra,~J.; Baboulin,~M. Towards dense linear algebra for hybrid GPU accelerated manycore systems. \emph{Parallel Computing} \textbf{2010}, \emph{36}, 232--240, Parallel Matrix Algorithms and Applications\relax
\mciteBstWouldAddEndPuncttrue
\mciteSetBstMidEndSepPunct{\mcitedefaultmidpunct}
{\mcitedefaultendpunct}{\mcitedefaultseppunct}\relax
\EndOfBibitem
\bibitem[Burke(2011)]{Burke2011}
Burke,~P.~G. \emph{R-Matrix Theory of Atomic Collisions}, 1st ed.; Springer Series on Atomic, Optical, and Plasma Physics; Springer Berlin Heidelberg: Berlin, Heidelberg, 2011; Vol.~61; pp XVIII + 746\relax
\mciteBstWouldAddEndPuncttrue
\mciteSetBstMidEndSepPunct{\mcitedefaultmidpunct}
{\mcitedefaultendpunct}{\mcitedefaultseppunct}\relax
\EndOfBibitem
\bibitem[Tennyson(2010)]{Tennyson2010}
Tennyson,~J. Electron–molecule collision calculations using the R-matrix method. \emph{Physics Reports} \textbf{2010}, \emph{491}, 29--76\relax
\mciteBstWouldAddEndPuncttrue
\mciteSetBstMidEndSepPunct{\mcitedefaultmidpunct}
{\mcitedefaultendpunct}{\mcitedefaultseppunct}\relax
\EndOfBibitem
\bibitem[Baye \latin{et~al.}(1998)Baye, Hesse, Sparenberg, and Vincke]{Baye1998}
Baye,~D.; Hesse,~M.; Sparenberg,~J.~M.; Vincke,~M. Analysis of the R-matrix method on Lagrange meshes. \emph{Journal of Physics B: Atomic, Molecular and Optical Physics} \textbf{1998}, \emph{31}, 3439--3454\relax
\mciteBstWouldAddEndPuncttrue
\mciteSetBstMidEndSepPunct{\mcitedefaultmidpunct}
{\mcitedefaultendpunct}{\mcitedefaultseppunct}\relax
\EndOfBibitem
\bibitem[Descouvemont and Baye(2010)Descouvemont, and Baye]{Descouvemont2010}
Descouvemont,~P.; Baye,~D. The R-matrix theory. \emph{Reports on Progress in Physics} \textbf{2010}, \emph{73}, 44\relax
\mciteBstWouldAddEndPuncttrue
\mciteSetBstMidEndSepPunct{\mcitedefaultmidpunct}
{\mcitedefaultendpunct}{\mcitedefaultseppunct}\relax
\EndOfBibitem
\bibitem[Bocchetta and Gerratt(1985)Bocchetta, and Gerratt]{Bocchetta1985}
Bocchetta,~C.~J.; Gerratt,~J. The application of the Wigner R-matrix theory to molecular collisions. \emph{The Journal of Chemical Physics} \textbf{1985}, \emph{82}, 1351--1362\relax
\mciteBstWouldAddEndPuncttrue
\mciteSetBstMidEndSepPunct{\mcitedefaultmidpunct}
{\mcitedefaultendpunct}{\mcitedefaultseppunct}\relax
\EndOfBibitem
\bibitem[Bocchetta \latin{et~al.}(1988)Bocchetta, Gerratt, and Guthrie]{Bocchetta1988}
Bocchetta,~C.~J.; Gerratt,~J.; Guthrie,~G. Study of He–H2CO collisions at interstellar temperatures using the L2 R‐matrix method. \emph{The Journal of Chemical Physics} \textbf{1988}, \emph{88}, 975--984\relax
\mciteBstWouldAddEndPuncttrue
\mciteSetBstMidEndSepPunct{\mcitedefaultmidpunct}
{\mcitedefaultendpunct}{\mcitedefaultseppunct}\relax
\EndOfBibitem
\bibitem[Tennyson \latin{et~al.}(2016)Tennyson, McKemmish, and Rivlin]{Tennyson2016}
Tennyson,~J.; McKemmish,~L.~K.; Rivlin,~T. Low-temperature chemistry using the R-matrix method. \emph{Faraday Discussions} \textbf{2016}, \emph{195}, 31--48\relax
\mciteBstWouldAddEndPuncttrue
\mciteSetBstMidEndSepPunct{\mcitedefaultmidpunct}
{\mcitedefaultendpunct}{\mcitedefaultseppunct}\relax
\EndOfBibitem
\bibitem[Rivlin \latin{et~al.}(2019)Rivlin, McKemmish, and Tennyson]{Rivlin_morse_rmat}
Rivlin,~T.; McKemmish,~L.~K.; Tennyson,~J. Low-Temperature Scattering with the R-Matrix Method: The Morse Potential. Quantum Collisions and Confinement of Atomic and Molecular Species, and Photons. Singapore, 2019; pp 257--273\relax
\mciteBstWouldAddEndPuncttrue
\mciteSetBstMidEndSepPunct{\mcitedefaultmidpunct}
{\mcitedefaultendpunct}{\mcitedefaultseppunct}\relax
\EndOfBibitem
\bibitem[Rivlin \latin{et~al.}(2019)Rivlin, McKemmish, Spinlove, and Tennyson]{Rivlin2019}
Rivlin,~T.; McKemmish,~L.~K.; Spinlove,~K.~E.; Tennyson,~J. Low temperature scattering with the R-matrix method: argon-argon scattering. \emph{Molecular Physics} \textbf{2019}, \emph{117}, 3158--3170\relax
\mciteBstWouldAddEndPuncttrue
\mciteSetBstMidEndSepPunct{\mcitedefaultmidpunct}
{\mcitedefaultendpunct}{\mcitedefaultseppunct}\relax
\EndOfBibitem
\bibitem[McKemmish and Tennyson(2019)McKemmish, and Tennyson]{McKemmish2019}
McKemmish,~L.~K.; Tennyson,~J. General mathematical formulation of scattering processes in atom-diatomic collisions in the RmatReact methodology. \emph{Philosophical Transactions of the Royal Society A: Mathematical, Physical and Engineering Sciences} \textbf{2019}, \emph{377}\relax
\mciteBstWouldAddEndPuncttrue
\mciteSetBstMidEndSepPunct{\mcitedefaultmidpunct}
{\mcitedefaultendpunct}{\mcitedefaultseppunct}\relax
\EndOfBibitem
\bibitem[Zare(1988)]{zare:88}
Zare,~R.~N. \emph{Angular Momentum}; Wiley: New York, 1988\relax
\mciteBstWouldAddEndPuncttrue
\mciteSetBstMidEndSepPunct{\mcitedefaultmidpunct}
{\mcitedefaultendpunct}{\mcitedefaultseppunct}\relax
\EndOfBibitem
\bibitem[García-Vázquez \latin{et~al.}(2023)García-Vázquez, Denis-Alpizar, and Stoecklin]{Garcia-Vazquez_2023}
García-Vázquez,~R.~M.; Denis-Alpizar,~O.; Stoecklin,~T. A Comparative Study of the Cold Collisions of H$_2$O and D$_2$O with Ne. \emph{The Journal of Physical Chemistry A} \textbf{2023}, \emph{127}, 4838--4847\relax
\mciteBstWouldAddEndPuncttrue
\mciteSetBstMidEndSepPunct{\mcitedefaultmidpunct}
{\mcitedefaultendpunct}{\mcitedefaultseppunct}\relax
\EndOfBibitem
\bibitem[García-Vázquez \latin{et~al.}(2024)García-Vázquez, Bergeat, Denis-Alpizar, Faure, Stoecklin, and Morales]{Garcia-Vazquez-2024-FD}
García-Vázquez,~R.~M.; Bergeat,~A.; Denis-Alpizar,~O.; Faure,~A.; Stoecklin,~T.; Morales,~S.~B. Scattering resonances in the rotational excitation of HDO by Ne and normal-H2: theory and experiment. \emph{Faraday Discussions} \textbf{2024}, \emph{251}, 205--224\relax
\mciteBstWouldAddEndPuncttrue
\mciteSetBstMidEndSepPunct{\mcitedefaultmidpunct}
{\mcitedefaultendpunct}{\mcitedefaultseppunct}\relax
\EndOfBibitem
\bibitem[Cabrera-González \latin{et~al.}(2022)Cabrera-González, Denis-Alpizar, Páez-Hernández, and Stoecklin]{lisan-h-h2o}
Cabrera-González,~L.~D.; Denis-Alpizar,~O.; Páez-Hernández,~D.; Stoecklin,~T. {Quantum study of the bending relaxation of H2O by collision with H}. \emph{Monthly Notices of the Royal Astronomical Society} \textbf{2022}, \emph{514}, 4426--4432\relax
\mciteBstWouldAddEndPuncttrue
\mciteSetBstMidEndSepPunct{\mcitedefaultmidpunct}
{\mcitedefaultendpunct}{\mcitedefaultseppunct}\relax
\EndOfBibitem
\bibitem[Yang \latin{et~al.}(2024)Yang, Qu, Bowman, Yang, Guo, Balakrishnan, Forrey, and Stancil]{H-H2O_6D}
Yang,~B.; Qu,~C.; Bowman,~J.~M.; Yang,~D.; Guo,~H.; Balakrishnan,~N.; Forrey,~R.~C.; Stancil,~P.~C. Inelastic Triatom-Atom Quantum Close-Coupling Dynamics in Full Dimensionality: All Rovibrational Mode Quenching of Water Due to the H Impact on a Six-Dimensional Potential Energy Surface. \emph{The Journal of Physical Chemistry Letters} \textbf{2024}, \emph{15}, 11312--11319, PMID: 39496299\relax
\mciteBstWouldAddEndPuncttrue
\mciteSetBstMidEndSepPunct{\mcitedefaultmidpunct}
{\mcitedefaultendpunct}{\mcitedefaultseppunct}\relax
\EndOfBibitem
\bibitem[Stoecklin \latin{et~al.}(2019)Stoecklin, Denis-Alpizar, Clergerie, Halvick, Faure, and Scribano]{Stoecklin_h2o-ph2_2019}
Stoecklin,~T.; Denis-Alpizar,~O.; Clergerie,~A.; Halvick,~P.; Faure,~A.; Scribano,~Y. Rigid-Bender Close-Coupling Treatment of the Inelastic Collisions of H2O with para-H2. \emph{The Journal of Physical Chemistry A} \textbf{2019}, \emph{123}, 5704--5712, PMID: 31192600\relax
\mciteBstWouldAddEndPuncttrue
\mciteSetBstMidEndSepPunct{\mcitedefaultmidpunct}
{\mcitedefaultendpunct}{\mcitedefaultseppunct}\relax
\EndOfBibitem
\bibitem[Stoecklin \latin{et~al.}(2021)Stoecklin, Cabrera-González, Denis-Alpizar, and Páez-Hernández]{Stoecklin_he-h2o}
Stoecklin,~T.; Cabrera-González,~L.~D.; Denis-Alpizar,~O.; Páez-Hernández,~D. A close coupling study of the bending relaxation of H2O by collision with He. \emph{The Journal of Chemical Physics} \textbf{2021}, \emph{154}, 144307\relax
\mciteBstWouldAddEndPuncttrue
\mciteSetBstMidEndSepPunct{\mcitedefaultmidpunct}
{\mcitedefaultendpunct}{\mcitedefaultseppunct}\relax
\EndOfBibitem
\bibitem[Baye(2015)]{BAYE20151}
Baye,~D. The Lagrange-mesh method. \emph{Physics Reports} \textbf{2015}, \emph{565}, 1--107\relax
\mciteBstWouldAddEndPuncttrue
\mciteSetBstMidEndSepPunct{\mcitedefaultmidpunct}
{\mcitedefaultendpunct}{\mcitedefaultseppunct}\relax
\EndOfBibitem
\bibitem[García-Vázquez \latin{et~al.}(2024)García-Vázquez, Cabrera-González, Denis-Alpizar, and Stoecklin]{Garcia_Vazquez_2024_ar}
García-Vázquez,~R.~M.; Cabrera-González,~L.~D.; Denis-Alpizar,~O.; Stoecklin,~T. A Rigid Bender Study of the Bending Relaxation of H2O and D2O by Collisions with Ar. \emph{ChemPhysChem} \textbf{2024}, \emph{25}, e202300752\relax
\mciteBstWouldAddEndPuncttrue
\mciteSetBstMidEndSepPunct{\mcitedefaultmidpunct}
{\mcitedefaultendpunct}{\mcitedefaultseppunct}\relax
\EndOfBibitem
\bibitem[Garc\'ia-V\'azquez \latin{et~al.}(2025)Garc\'ia-V\'azquez, Stoecklin, and Halvick]{GarciaVazquez2025}
Garc\'ia-V\'azquez,~R.~M.; Stoecklin,~T.; Halvick,~P. manuscript submitted to \textit{Physical Review A}\relax
\mciteBstWouldAddEndPuncttrue
\mciteSetBstMidEndSepPunct{\mcitedefaultmidpunct}
{\mcitedefaultendpunct}{\mcitedefaultseppunct}\relax
\EndOfBibitem
\bibitem[{IDRIS, CNRS}(2024)]{jeanzayGPU}
{IDRIS, CNRS} Jean Zay Supercomputer--GPU Partition Specifications. \url{http://www.idris.fr/eng/jean-zay/cpu/jean-zay-cpu-hw-eng.html#gpu_p6}, 2024; Accessed: 2025-10-20\relax
\mciteBstWouldAddEndPuncttrue
\mciteSetBstMidEndSepPunct{\mcitedefaultmidpunct}
{\mcitedefaultendpunct}{\mcitedefaultseppunct}\relax
\EndOfBibitem
\end{mcitethebibliography}

\end{document}